\documentclass[12pt]{article}

\pdfoutput=1

\usepackage{jcappub}
\usepackage{bm}
\usepackage{amssymb,amsmath,amsthm}
\usepackage{graphics}

\def\widebar{\overline}

\begin{document}
\title{Attracted to de Sitter:\\ cosmology of the linear Horndeski models}
\author{Prado Mart\'{\i}n-Moruno,  Nelson J.~Nunes {\rm and} Francisco S.~N.~Lobo}
\affiliation{Instituto de Astrof\'isica e Ci\^encias do Espa\c{c}o, 
Faculdade de Ci\^encias da Universidade de Lisboa, Campo Grande, PT1749-016 Lisboa, Portugal}
\emailAdd{pmmoruno@fc.ul.pt}
\emailAdd{njnunes@fc.ul.pt}
\emailAdd{fslobo@fc.ul.pt}

\abstract{
We consider Horndeski cosmological models, with a minisuperspace Lagrangian linear in the field derivative, that are able to screen {\it any} vacuum energy and material content leading to a spatially flat de Sitter vacuum fixed by the theory itself. Furthermore, we investigate particular models with a cosmic evolution independent of the material content and use them to understand the general characteristics of this framework. We also consider more realistic models, which we denote the ``term-by-term'' and ``tripod'' models, focusing attention on cases in which the critical point is indeed an attractor solution and the cosmological history is of particular interest.

\bigskip
\noindent
\today
}

\maketitle


\section{Introduction}

Scalar-tensor theories of gravity have recently received a renewed interest in relation with the problem of describing the current accelerated expansion of our Universe. Indeed, scalar fields are ubiquitous 
in theories of high energy physics beyond the standard model and, in particular, are present in theories which include extra spatial dimensions, such as those derived from string theories. The initial simplest model, denoted by the quintessence field \cite{Zlatev:1998tr}, consisted of a minimally coupled single scalar field self-interacting through a scalar potential, and with canonical kinetic terms. An interesting model, described as coupled quintessence, which analysed the consequences of coupling the field responsible for acceleration to matter fields \cite{Amendola:1999er}, was rapidly extended to consider non-canonical kinetic terms, with higher powers of the field's velocity, denoted $k-$essence \cite{ArmendarizPicon:2000ah}, and non-minimal couplings to gravity. Scalar fields with Lagrangians presenting nonlinear kinetic interactions and possessing shift symmetries, known as galileons, were found to appear in the decoupling limit of the DGP models \cite{Dvali:2000hr}. However, the ``self-accelerating'' solution in the DGP model is plagued by a ghost instability, rendering the solution untenable. This fact motivated the proposal of an infrared modification of General Relativity (GR) \cite{Nicolis:2008in}, involving an internal ``galilean'' invariance, where the symmetry constrained the structure of the effective Lagrangian without introducing ghosts. 
These galileon fields have second order equations of motion and lead to the Vainshtein screening mechanism at astrophysical scales \cite{deRham:2012az}. Galileon scalar fields also appear in the decoupling limit of 
the ghost-free theory of massive gravity \cite{deRham:2010ik, deRham:2010tw}. 
The understanding of this theory \cite{deRham:2010kj,Hassan:2011tf} as well as of 
the intimately related \cite{Baccetti:2012bk} bigravity theory \cite{Hassan:2011zd}, free from the Boulware--Desser ghost \cite{BD} has also recently been extensively explored 
(we refer the reader to Ref.~\cite{deRham:2014zqa} for a recent review).

In fact, given the large number of models, the question arises how one should study and compare them in a unified manner, 
and determine which if any is the origin of cosmic acceleration. 
A particularly useful tool in this direction is the realisation that the covariantized Galileons \cite{Deffayet:2009wt}
(recovered in Ref.~\cite{deRham:2010eu})
are special cases of the most general Lagrangian which leads to second order field equations \cite{Deffayet:2011gz}, 
and therefore yielding the model free of Ostrogradski instabilities. 
(This is, of course, also the case for the covariantization of the decoupling limit of massive gravity presented in 
Refs.~\cite{Chkareuli:2011te, deRham:2011by}.)
This Lagrangian was first written down by Horndeski in 1974 \cite{Horndeski:1974wa} and recently rediscovered by Deffayet 
{\it et al.}~in 2011 \cite{Deffayet:2011gz} in an equivalent formulation \cite{Kobayashi:2011nu}. 
An appealing theoretical property is the possibility that the field is able to self-tune considering some
terms within the Horndeski Lagrangian (see Refs.~\cite{Dvali:2007kt,deRham:2010tw} for early works on degravitation). 
This can be used to partially alleviate the cosmological constant problem, which is not only based on the huge fine-tuning needed 
to cancel the value of the vacuum energy in particle physics
to produce the value of the cosmological constant as inferred from cosmological observations, 
but also on the need to fine-tune again this difference after any phase transition.
In particular these terms have been investigated with the aim of finding a viable self-tuning mechanism, which screens the spacetime curvature from the net cosmological constant \cite{deRham:2014zqa}. 
Horndeski models are  able to self-tune any vacuum energy leading to a Minkowski vacuum space have, therefore, 
been obtained \cite{Charmousis:2011bf,Charmousis:2011ea}.

Starting from the most general scalar-tensor theory with second order field equations in four dimensions, the unique action 
that allows for the existence of a consistent self-tuning mechanism on FLRW backgrounds was established in Ref.~\cite{Charmousis:2011bf}. 
It consists of a combination of only four base Lagrangians, denoted the {\it Fab Four}, with an intriguing geometric structure dependent 
on the Ricci scalar, the Einstein tensor, the double dual of the Riemann tensor and the Gauss-Bonnet combination. 
It was shown that spacetime curvature can be screened from the net cosmological constant at any given moment, 
as the scalar field is allowed to break Poincar\'e invariance on the self-tuning vacua, 
thereby evading the Weinberg no-go theorem \cite{Weinberg:1988cp}.
This is only possible, however, when there is a nonvanishing spatial curvature, a requirement that has raised instabilities
related concerns in the literature \cite{deRham:2011by}.
The cosmology of the Fab-Four Lagrangian, which also screens the contribution of any matter content, has been studied in Ref.~\cite{Copeland:2012qf}. By performing a phase plane analysis of the system, a number of fixed points for the system were obtained. In addition to obtaining inflationary solutions, conventional radiation/matter-like solutions were also found, but in regimes where the energy density is dominated by a cosmological constant, without the explicit forms of radiation or matter. An extended period of `matter domination' was shown to exist, which opens up the possibility that cosmological structures can be generated, and thus recovering a consistent cosmology 
before the Minkowski attractor is reached.
While not yet a full theory of cosmological evolution, such a mechanism is intriguing, and leads to the question of whether realistic cosmologies that  include this mechanism as well as the late time acceleration, can be constructed.

Some of the terms appearing in the Fab Four models \cite{Charmousis:2011bf,Charmousis:2011ea} have been generalized \cite{Appleby:2012rx}. More specifically, it was shown that a nonlinear combination of 
purely kinetic gravity terms can give rise to an accelerating universe without the addition of extra propagating degrees of freedom on cosmological backgrounds, and exhibit self tuning to bring a large cosmological constant under control. However, concerns of the viability of the cosmologies described by this generalization have been raised in Ref. \cite{Linder:2013zoa}. Indeed, these generalizations may be ghost free and stable, have attractor solutions in the past and future, and possess self tuning that solves the original cosmological constant problem, however, it was shown that they do not possess all of these qualities at the same time. 
It was also demonstrated that the screening is so powerful that it not only cancels the cosmological constant but also all other energy densities.
Nevertheless, it must be noted that this is a general characteristic of self-tuning models, and it only entails a problem for models in which the self-tuning takes place so fast that it 
preclude the existence of a matter dominated phase before the accelerated expansion.

More recently, Horndeski models which can screen any vacuum energy to a spatial flat de Sitter vacuum fixed by the theory itself have been investigated \cite{Nosotros}.
Thus, the most general minisuperspace Lagrangian for a scalar field leading to an equation of motion without higher than second derivatives and a spatially flat de Sitter critical point has been presented \cite{Nosotros}.
These models have the advantage of allowing to understand the current epoch of accelerating expansion as the result 
of the approach to a stable critical point while avoiding the cosmological constant problem.
In this paper we investigate the cosmology of a family of these models which have a minisuperspace Lagrangian linear on the field derivatives.

This article is outlined in the following manner: In Section \ref{model}, we present the model and include some expressions that will later be useful for our study. In Section \ref{dynamical}, we present some considerations regarding the dynamical problem and conclude that one needs to go beyond the linear stability analysis to extract any conclusion about the stability of the critical point, which can be done by considering specific models. In Section \ref{nomatter}, we include some interesting models for which the dynamics of the universe is completely independent of the material content. This study is of great relevance for understanding the dynamical evolution of more complicated models in the later sections. In Section \ref{simplelinearcosmology}, we consider a subfamily of models, which we denote the ``term-by-term'' models for reasons that will become apparent along the article. As we will show these models can describe a matter dominated universe followed by a de Sitter expansion. In Section \ref{tripod}, we consider what we call the ``tripod'' models, which are based on three potentials. These models can be the simplest examples of models belonging to this family which reproduce a consistent cosmological history. Finally, in Section \ref{discussion}, we summarize and discuss our results.

\section{The model}\label{model}

The most general Horndeski models with a spatially flat de Sitter critical point for any material content can be classified in two families \cite{Nosotros}. Their main characteristic is that the first family
has a point-like Lagrangian linear on the field derivatives and the second family has a minisuperspace Lagrangian which vanishes on shell in $a$, that is, when it is evaluated at the scale factor of the critical point.
The point-like Lagrangian for these two families are written as \cite{Nosotros}
\begin{equation}\label{Ltot}
 L=L_{\rm EH}+L_{\rm linear}+L_{\rm nonlinear}+L_{\rm m},
\end{equation}
where
\begin{equation}\label{LEH}
 L_{\rm EH}=-3M_{\rm Pl}^2\,a^3\,H^2,
\end{equation}
is the usual Einstein-Hilbert Lagrangian with $H$ the Hubble parameter,
\begin{equation}\label{Llinearg}
 L_{\rm linear}=a^3\sum_{i=0}^{3}\left[3\sqrt{\Lambda} \,U_i(\phi)+\dot\phi\,W_i(\phi)\right]\,H^i,
\end{equation}
is the Lagrangian of the first family of models, where $U_i$ and $W_i$ being functions of $\phi$, and
\begin{equation}\label{Lnonlinear}
L_{\rm nonlinear}=a^3 \sum_{i=0}^{3}X_i\left(\phi,\,\dot\phi\right)H^i,
\end{equation}
describes the second family, with $X_i$ being arbitrary functions.  In addition, 
\begin{equation}\label{Lm}
 L_{\rm m}= -a^3\rho(a)=-a^3\sum_s\rho_{s}(a),\qquad {\rm with}\qquad \rho_{s}(a)=\rho_{s,0}a^{-3(1+w_s)},
\end{equation}
is the Lagrangian for the minimally coupled matter species labeled by $s$, and the conditions
\begin{equation}\label{condition}
  \sum_{i=0}^{3}W_i(\phi)\Lambda^{i/2}=\sum_{j=0}^{3}U_{j,\phi}(\phi)\Lambda^{j/2},
\end{equation}
and
\begin{equation}\label{condition2}
 \sum_{i=0}^{3} X_i\left(\phi,\,\dot\phi\right)\Lambda^{i/2}=0,
\end{equation}
have to be satisfied. Conditions (\ref{condition}) and (\ref{condition2}) 
guarantee that the Lagrangian evaluated at the critical point is linear on $\dot\phi$ which, at the end of the day, is
  the condition needed for the existence of a spatially flat de Sitter critical point for any material content (see \cite{Nosotros}
for technical details).

In this article, we focus our attention on the first family of models. Therefore, we consider that the total Lagrangian is given by
\begin{equation}\label{Ltot2}
 L=L_{\rm EH}+L_{\rm linear}+L_{\rm m}.
\end{equation}
The field equation for these models is given by \cite{Nosotros}
\begin{equation}\label{field}
 3\sqrt{\Lambda}\sum_{i=0}^{3}H^i\left[U_{i,\phi}(\phi)-\frac{H}{\sqrt{\Lambda}}W_i(\phi)\right]=\dot H\sum_{i=0}^{3}iH^{i-1} W_i(\phi).
\end{equation}
Thus, we avoid models for which the left hand side of Eq. (\ref{field}) vanishes,
otherwise, there is no dynamical evolution of $H$, as $\dot{H}$ would also have to vanish.

The modified Friedmann equation can be expressed using the Hamiltonian density which is
\begin{equation}\label{Htodo}
 \mathcal{H}=\mathcal{H}_{\rm EH}+\mathcal{H}_{\rm linear}+\mathcal{H}_{\rm m}=0,
\end{equation}
where the terms are given by
\begin{eqnarray}
\label{HEH}
 \mathcal{H}_{\rm EH}&=&-3M_{\rm Pl}^2H^2, \\
\label{HL}
\mathcal{H}_{\rm linear}&=&\sum_{i=0}^{3} \left[3(i-1)\sqrt{\Lambda}\,U_i(\phi)+i\,\dot\phi\,W_i(\phi)\right]H^i\equiv\rho_\phi, \\
\label{Hm}
 \mathcal{H}_{\rm m}&=&\rho(a),
\end{eqnarray}
respectively. We have defined $\rho_\phi$ for obvious reasons.
It must be noted that for the model to be able to screen by itself, 
the Hamiltonian density has to contain at least one term with a function of $\dot\phi$.
Thus, we must have at least one $W_i\neq0$ with $i\neq0$ for a dynamical approach to a de Sitter vacuum.
The modified Friedmann equation (\ref{Htodo}) can be explicitly written as
\begin{equation}\label{Friedmann}
 H^2=\frac{1}{3M_{\rm Pl}^2}\left(\rho+\rho_\phi\right).
\end{equation}
Defining $\Omega=8\pi G\rho/(3H^2)$ and $\Omega_\phi=8\pi G\rho_\phi/(3H^2)$, with $\rho_\phi$ given by Eq.~(\ref{HL}), then Eq. (\ref{Friedmann}) can be expressed as
\begin{equation}\label{Friedlinear}
 \Omega+\Omega_\phi=1.
\end{equation}
It is also useful to define a dark energy equation of state parameter as
\begin{equation}\label{wphi}
 w_\phi=-1-\frac{1}{3H}\frac{\dot\rho_\phi}{\rho_\phi}, 
\end{equation}
which should not be confused with the effective equation of state parameter
\begin{equation}\label{weff}
 w_{\rm eff}=-1-\frac{2\dot H}{3H^2}.
\end{equation}

\section{Dynamical system}\label{dynamical}

Defining $N=\ln a$, the field equation (\ref{field}) and the modified Friedmann equation (\ref{Friedmann}) are written as
\begin{eqnarray}
\label{H'}
 H'&=&3\frac{\sum_{i=0}^{3}H^i\left[\sqrt{\Lambda}\,\widebar U_{i,\phi}(\phi)-H\,\widebar W_{i}(\phi)\right]}{\sum_{i=0}^3i\,H^{i}\widebar W_i(\phi)}, \\
\label{phi'}
 \phi'&=&\sqrt{\Lambda}\frac{\left(1-\Omega\right)H^2-3\sum_{i=0}^{3}(i-1)\,H^i\,\widebar U_i(\phi)}{\sum_{i=0}^{3}i\,H^{i+1}\widebar W_i(\phi)},
\end{eqnarray}
respectively, 
subject to the constraint
\begin{equation}
\sum_{i=0}^3 \widebar W_i \Lambda^{i/2} = 
\sum_{j=0}^3 \widebar U_{j,\phi} \Lambda^{j/2},
\end{equation}
where we have defined $\widebar U_i=8\pi G \sqrt{\Lambda}\,U_i/3$ and $\widebar W_i=8\pi G \sqrt{\Lambda}\,W_i/3$ for notational simplicity.
In order to close the autonomous system we assume that the different matter species can be treated as noninteracting, then considering $\Omega=\sum_s\Omega_s$ we have
\begin{equation}\label{m'}
\Omega_{s}'=-3 \Omega_{s}\left[1+w_s+2\frac{\sum_{i=0}^{3}H^i\left[\sqrt{\Lambda}\,\widebar U_{i,\phi}(\phi)-H\,\widebar W_{i}(\phi)\right]}{\sum_{i=1..3}i\,H^{i+1}\widebar W_i(\phi)}\right],
\end{equation}
which simply arises from the conservation equation, already integrated in Eq. (\ref{Lm}) and equivalent to
\begin{equation}\label{Omegaint}
\Omega_{s}=\frac{\Omega_{s,0}H_0^2}{H^2}\,{\rm exp}\left[-3(1+w_s)N\right],
\end{equation}
through
\begin{equation}\label{eq3}
\Omega_{s}'=-\Omega_{s}\left[3(1+w_s)+2\frac{H'}{H}\right].
\end{equation}
In the first place, taking into account condition (\ref{condition}) one can note that Eq.~(\ref{H'}) trivially vanishes for $H=\sqrt{\Lambda}$, as it should by construction. Thus, there is a de Sitter critical point with $H_{\rm c}=\sqrt{\Lambda}$ for any material content. 
In the second place, from Eq.~(\ref{eq3}), one can note that a critical point which exists for any material content (that is, any value of $w_i$) has to be characterized by $\Omega_{s,c}=0$, where the 
subscript c means evaluation at the critical point.
In third place, considering also Eq.~(\ref{phi'}) one concludes that we have a de Sitter critical point given by
\begin{equation}\label{phic}
 \{H_{\rm c}=\sqrt{\Lambda},\,\phi=\phi_c,\,\Omega_{s,{\rm c}}=0\},\,\,\,\, {\rm with}\,\,\,\, \phi_c\,\,\,\,{\rm such\,\,as}\,\,\,\, \sum_{i=0}^{3}(i-1)\,\Lambda^{i/2}\,\widebar U_i(\phi_c)=\frac{\Lambda}{3}.
\end{equation}
The last relation comes from demanding $\phi'=\Omega=0$, $\phi=\phi_{\rm c}$ and $H = \sqrt{\Lambda}$ in 
Eq.~(\ref{phi'}). 
Although the existence of this de Sitter critical point is independent of the particular form of the potentials, its position
in the phase space depends on them. Thus, if one wants the cosmological model to be well-defined close to the critical point, 
one has to require $\phi_c$ to be real. 
If for a particular model there is more than one real solution $\phi_c$ to Eq.~(\ref{phic}), 
then we have various de Sitter critical points characterized by $H_{\rm c}=\sqrt{\Lambda}$ and $\Omega_{s,{\rm c}}=0$ 
(but with different $\phi_c$'s).
On the other hand, the existence of the critical point (or points) given by Eq.~(\ref{phic}) is not precluding the potential 
existence of additional critical points. Thus, in particular models there could be additional critical points of the dynamics apart 
from our de Sitter critical point (or points).

Assuming now only one material component and considering linear stability theory, 
the eigenvalues of the Jacobian matrix of the system given by Eqs. (\ref{H'})--(\ref{m'}) at the critical point (\ref{phic})
are
\begin{equation}
 \lambda=0,\hspace{1cm} \lambda=-3,\hspace{1cm}  \lambda=-3(1+w_s),
\end{equation}
where we have simplified the resulting expression considering condition (\ref{condition}). 
Therefore, in order to conclude whether this critical point is stable or not, one should go beyond linear order.
This is not a trivial undertaking without specifying the particular potentials.
Nevertheless, we can consider the 2-dimensional dynamical problems defined in the sections of the phase space given by the critical values to look for some hints about the behaviour of the 3-dimensional system.
Firstly, if one chooses to restrict attention to a vacuum model, i.e. $\Omega_{s,c}=0$, one also obtains that one of the two eigenvalues vanishes. 
Secondly, restricting attention to the 2-dimensional dynamical problem defined in the section  of the phase space $H_{\rm c}=\sqrt{\Lambda}$ one obtains again one eigenvalue $\lambda=0$. 
Thirdly, in the section given by $\phi=\phi_c$, the 2-dimensional Jacobian matrix evaluated at the critical point has non-vanishing eigenvalues, 
one for $\lambda=-3(1+w_s)$ and the sign of the other depending on the particular potentials. 
Thus, for the 2-dimensional dynamical system defined in this section, if the critical point 
is an attractor for a model with a given value of $w_s$ such that $1+w_s>0$, it is also an attractor for a different value of $w_s>-1$, whereas it is unstable for models with $w_s<-1$.
As we will see studying particular examples, this dependence of the stability on the sign of $1+w_s$ is present  in some 3-dimensional models.

\section{Models in which matter does not matter}\label{nomatter}

Reflecting on the form of the equations of the dynamical system, Eqs.~ (\ref{H'})--(\ref{m'}), one notes that the equation for $H'$ is independent of the material content. Nevertheless, in order
to integrate this equation to obtain $H(N)$ one must substitute the expression for $\phi(N)$, which depends on $\Omega_{\rm s}$ through Eq.~(\ref{phi'}). Thus, even if there is no direct dependence on
$\Omega_{\rm s}$ on Eq.~(\ref{H'}), the dynamics of the universe generically depends on the material content which enters into play through the function $\phi(N)$. Nevertheless, it can be seen
that for some models the dependence of Eq.~(\ref{H'})  on $\phi$  can disappear and, therefore, the evolution of the universe is independent of the matter content for those models. 

In this section, we consider some of these models for which any cosmological constant or material is completely screened by the field contribution during the whole cosmological history 
even if the self tuning to de Sitter can take place dynamically.
Firstly, we will show why the $W_i$ potential present in the model cannot be $W_0$ if we want to approach a de Sitter solution dynamically.
Secondly, we will consider models with only one $W_i$ and one $U_j$. Thirdly, as we need to have at least one potential $W_j$ with $j\neq0$ but it is not necessary to have any $U_{i}$, 
we will include some comments about models with two $W_i$ potentials.
Finally, we discuss an additional case for which the evolution of $H$ is independent of the material content.

\subsection{$W_0$: exact de Sitter evolution}

If the only $W_i\neq0$ is $W_0$, then $\mathcal{H}_{,\dot\phi}=0$ from Eq.~(\ref{HL}). The model is not able to screen dynamically and it has just an exact de Sitter evolution
(incompatible with a consistent cosmic history) independently of the number of $U_i$'s present. 
We can understand this better by noting that condition (\ref{condition}) and the field equation (\ref{field}) imply
\begin{equation}
 W_0=\sum_{i=0}^{3}U_{i,\phi}\Lambda^{i/2}
\end{equation}
and
\begin{equation}
 W_0=\frac{\sqrt{\Lambda}}{H}\sum_{i=0}^{3}U_{i,\phi}H^{i},
\end{equation}
respectively. These two equations can only be simultaneously satisfied if $H=\sqrt{\Lambda}$ during the whole evolution.

\subsection{One $W_i$, $U_j$ pair}\label{couple}

Assuming that we have only one $W_i$ and one $U_j$, with $i\neq0$, condition (\ref{condition}) implies
\begin{equation}
 W_i=\Lambda^{\frac{j-i}{2}}\,U_{j,\phi},
\end{equation}
which gives the same relation between $\widebar W_i$ and $\widebar U_{j,\phi}$.
Thus, Eq.~(\ref{H'}) can be written as
\begin{equation}\label{hinde}
 h'=\frac{3\,h}{i}\left(h^n-1\right),\qquad{\rm with}\qquad n=j-i-1\neq0,
\end{equation}
where $h=H/\sqrt{\Lambda}$. This differential equation is independent of $\phi$ or $\Omega_{s}$, therefore, one can integrate $h'$ and obtain the dynamics of the universe without any knowledge of the material 
content or the particular form of the potentials.
It should be emphasized that if $n=0$, then we are in a case given by Eq.~(\ref{condition2}) and, thus, $h'=0$ during the whole evolution. 
Furthermore, one can integrate Eq.~(\ref{hinde}) to obtain
\begin{equation}
 \mid 1-h^{-n}\mid=D\,{\rm exp}\left(\frac{3n}{i}N\right).
\end{equation}
which implies that for $h>1$ and $n>0$ and for $h<1$ and $n<0$ we have
\begin{equation}\label{hs1}
 h=\left[1-D\,{\rm exp}\left(\frac{3n}{i}N\right)\right]^{-1/n},
\end{equation}
whereas for  $h<1$ and $n>0$ and for $h>1$ and $n<0$ we have
\begin{equation}\label{hs2}
 h=\left[1+D\,{\rm exp}\left(\frac{3n}{i}N\right)\right]^{-1/n},
\end{equation}
where $D>0$ is an integration constant.
Therefore, for $n>0$ one has that the de Sitter critical point is a repeller, since for one branch of solutions $h$ evolves from $1$ at $N\rightarrow-\infty$ to infinity at a finite $N$,
and for the other branch, to $h\rightarrow0$ when $N\rightarrow\infty$. 
On the other hand, for $n<0$ we have that the de Sitter critical point is an attractor. This attractor is approached from values of $h$ larger than $1$ for solutions described by
Eq. (\ref{hs2}), and smaller for those given by Eq. (\ref{hs1}). 

Let us now focus our attention on the models with a de Sitter critical point, that is, modes with $n<0$, and consider the branch in which $h$ is a decreasing function of $N$.
In this case, considering the limit $h\gg1$ of the r.~h.~s.~of Eq.~(\ref{hinde}), we obtain
\begin{equation}
h'\simeq-\frac{3}{i}h\qquad {\rm for}\qquad h\gg1,
\end{equation}
for any negative $n$.
Thus, far from the attractor the models behave as having
\begin{eqnarray}
\label{smc1}
&& w_{\rm eff}\simeq1\,\, {\rm if\,\,} i=1:\,\, {\rm (stiff\,matter)},\\
 \label{smc2}
 1+w_{\rm eff}\simeq\frac{2}{i}\qquad &\Rightarrow \qquad &  w_{\rm eff}\simeq0 \,\,{\rm if\,\,} i=2:\,\,{\rm (dust)},
 \\
 \label{smc3} &&  w_{\rm eff}\simeq\frac{2}{3} \,\,{\rm if\,\,} i=3:\,\,{\rm (curvature)}.
\end{eqnarray}
Moreover, considering a particular value of $n$, one can integrate the exact equations for $h'$, Eqs.~(\ref{hs1}) and (\ref{hs2}), to obtain the time evolution of the scale factor. For example,
for $n=-1$ (which is related to some considerations that we will include in  Sec. \ref{simplelinearcosmology}), we have
\begin{equation}\label{asimple}
 a(t)=\left[\pm D+C\,{\rm exp}\left(\frac{3\sqrt{\Lambda}}{i}t\right)\right]^{i/3},
\end{equation}
where the $\pm$ signs take into account both branches and $C>0$.

On the other hand, even if some models can reproduce a matter dominated epoch at early times, it should be emphasized that the material content is not affecting the cosmological dynamics at all.
Taking into account Eq. (\ref{Friedlinear}), one can see that the contribution of the material content is completely screened by the field through the
evolution. We have
\begin{equation}
 \Omega=1-3(j-1)\,\widebar U_j\,H^{j-2}+i\phi'\sqrt{\Lambda}\, \widebar U_{j,\phi}\,H^{i-1}.
\end{equation}
The field only appears in this equation but not on Eq.~(\ref{hinde}), hence, it has no effect on the evolution of the Universe.

\subsection{One  $W_i,\,W_j$ pair}\label{couple2}

If we consider a model with only two $W_i$ potentials and vanishing or constant $U_i$'s, condition (\ref{condition}) implies that we have
\begin{equation}
 W_i=-\Lambda^{\frac{j-i}{2}}\,W_j,
\end{equation}
which $j\neq i$ in order to obtain dynamics, as pointed out after Eq.~(\ref{field}).
In this case Eq.~(\ref{H'}) can be simplified to
\begin{equation}\label{h'W2}
 h'=-3h\,\frac{1-h^{i-j}}{j-ih^{i-j}}.
\end{equation}
Thus, one obtains
\begin{equation}
 \mid h^j\left(1-h^{i-j}\right)\mid= D e^{-3N},
\end{equation}
which approaches $0$ when $N\rightarrow\infty$, implying that either the first factor vanishes (being $h$ zero or infinity large depending on the sign of $j$) or $h\rightarrow 1$. Thus,
at least for some of these models, the de Sitter critical point is an attractor.

Considering the regime $h\gg1$ in Eq. (\ref{h'W2}) one gets
\begin{equation}
 h'\simeq-\frac{3}{j}h\qquad{\rm for}\qquad i<j,
\end{equation}
and
\begin{equation}
 h'\simeq-\frac{3}{i}h\qquad{\rm for}\qquad i>j.
\end{equation}
Thus, for the models which may describe a universe with a Hubble parameter decreasing towards the value of the attractor, one could mimic a cosmological phase dominated by stiff matter ($i=1$), dust ($i=2$),
or curvature ($i=3$) at early times, as explicitly shown in Eqs.~(\ref{smc1})--(\ref{smc3}). As in the case of Sec. \ref{couple}, the material content is perfectly screened by the field during the whole cosmic evolution.

\subsection{Particular cases}

Let us consider models with
\begin{equation}
 W_i=\Lambda^{-i/2}\beta_i,\qquad{\rm and}\qquad U_{i,\phi}=\Lambda^{-i/2}\alpha_i,
\end{equation}
with $\alpha_i$ and $\beta_i$ constant parameters and the factors $\Lambda$ have been included for convenience. Thus, condition (\ref{condition}) implies
$\sum_{i=0}^{3}\alpha_i=\sum_{i=0}^{3}\beta_i$.
In this case Eq.~(\ref{H'}) is again independent of $\phi$ and, therefore, on the material content. Indeed we have
\begin{equation}
 h'=3\frac{\sum_{i=0}^{3}h^i\left(\alpha_i-h\beta_i\right)}{\sum_{i=0}^{3}ih^i\beta_i},
\end{equation}
which can be integrated specifying the value of the parameters. Nevertheless, it should be noted that this case is genuinely different from those previously shown, since now $h'$ depends only on $h$ once we have
substituted the particular form of the potentials and not just considering which are different from zero.
Moreover, one expects that other particular choices of $W_i$ and $U_i$ can also lead to a differential equation for $h$ independent of $\phi$ and, therefore, a cosmological evolution independent of the matter
content.

\section{Term-by-Term models (TBT)}
\label{simplelinearcosmology}

\subsection{General considerations}

In this section, we investigate a particular family of models whose evolution depends on the matter content and, at the same time, can be studied in some generality without restricting the particular form of the potentials from the beginning. These models are given by assuming that condition (\ref{condition}) is satisfied term by term, ``Term-by-Term'' models. That is, we have
\begin{equation}
 W_i(\phi)=U_{i,\phi}(\phi)=\Lambda^{-i/2} V_{i,\phi}(\phi),
\end{equation}
and, thus, we are left with just four independent potentials.
Defining
\begin{equation}
 \widebar V_i(\phi)=\frac{8\pi\,G}{3\sqrt{\Lambda}}V_i(\phi),
\end{equation}
and again $h=H/\sqrt{\Lambda}$, we can write the field equation and the modified Friedmann equation, expressed as in Eqs.~(\ref{H'}) and (\ref{phi'}), as
\begin{eqnarray}
\label{hs'}
 h'&=& 3(1-h)\frac{\sum_{i=0}^{3}h^{i}\widebar V_{i,\phi}}{\sum_{i=0}^{3}i\,h^{i}\widebar V_{i,\phi}}, \\
\label{phis'}
 \phi'&=&\frac{\left(1-\Omega\right)h^2-3\sum_{i=0}^{3}(i-1)\widebar V_{i}h^i}{\sum_{i=0}^{3}i\,\widebar V_{i,\phi}h^{i+1}},
\end{eqnarray}
respectively. As in the general case, these equations together with the conservation of matter, i.e., Eq. (\ref{m'}), form an autonomous closed system.
Therefore, there is a critical point for any value of $w_s$ and number of species $s$ at $ \{h_c=1,\,\phi_c,\,\Omega_{s,{\rm c}}=0\}$,
where $\phi_c$ is such that
\begin{equation}\label{phics}
 \widebar V_0(\phi_c)=\widebar V_2(\phi_c)+2\widebar V_3(\phi_c)-\frac{1}{3}.
\end{equation}
As we have already pointed out for the general case, this critical point is not hyperbolic and, thus, the linear analysis is not enough to extract any conclusion about its stability. Thus, one has to consider particular models to conclude whether this point is stable or not.

In order to consider a particular model which could describe the current evolution of our Universe, let us reflect on the form of Eqs. (\ref{hs'}) and (\ref{phis'}).
As we have already emphasized in Sec. \ref{nomatter}, for matter to affect the dynamics one needs to have a differential equation for $h'(N)$ which depends on $\phi(N)$, with $\phi(N)$ given through Eq. (\ref{phis'})
in terms of $\Omega(N)$. This requires the presence of
at least two potentials with a different functional form and with non-vanishing derivatives. Moreover, in order to have self-tuning we need to have at least one $\widebar V_{i,\phi}\neq0$ for $i\neq0$.
On the other hand, one can study how  the dynamics of the models in the regions where a particular potential dominates in Eq.~(\ref{hs'}) taking into account considerations similar to those
presented for the models in Sec. \ref{couple}. In the present case, however, matter affects the dynamics, since the realization of the different regions of domination as well as their duration 
depend on the particular function $\phi(N)$ and, therefore, on $\Omega(N)$.
Nevertheless, this approach can suggest which models are candidates to describe a consistent background cosmological 
history.
When a given potential $\widebar V_i$ dominates we would have an expansion similar to that given by Eq.~(\ref{hinde}) for $n=-1$, namely,
\begin{equation}
 h'\simeq-\frac{3}{i}(h-1)\qquad\Rightarrow\qquad  h'\simeq-\frac{3}{i}h\qquad {\rm for}\qquad h\gg1.
\end{equation}
It must be noted that $\widebar V_0$ can never dominate the denominator of Eq. (\ref{hs'}) since it is not present there. If the critical point is an attractor and $h(N)$ decreases  towards this attractor, $h=1$,
we have a dynamics similar to general relativity for early times with
\begin{eqnarray}\label{weff2}
&& w_{\rm eff}\simeq1\,\, {\rm if\,\,} i=1:\,\,{\rm (stiff\,matter)},\\
 1+w_{\rm eff}\simeq\frac{2}{i}&\Rightarrow&  w_{\rm eff}\simeq0 \,\,{\rm if\,\,} i=2:\,\,{\rm (dust)},
 \\ &&  w_{\rm eff}\simeq\frac{2}{3} \,\,{\rm if\,\,} i=3:\,\,{\rm (curvature)},
\end{eqnarray}
as in the models of Sec. \ref{couple}.
Thus, if we want a model compatible with a matter dominated phase before accelerated expansion, we should investigate models with the potential $\widebar V_2$ and at least one additional potential.
The duration of the cosmological epochs as well as their order depend, however, of the material content and the particular form of the potentials through Eqs. (\ref{hs'}) and (\ref{phis'}).

\subsection{Matter dominated universe with a de Sitter attractor}

\begin{figure}[h]
\centering
\includegraphics[width=0.5\textwidth]{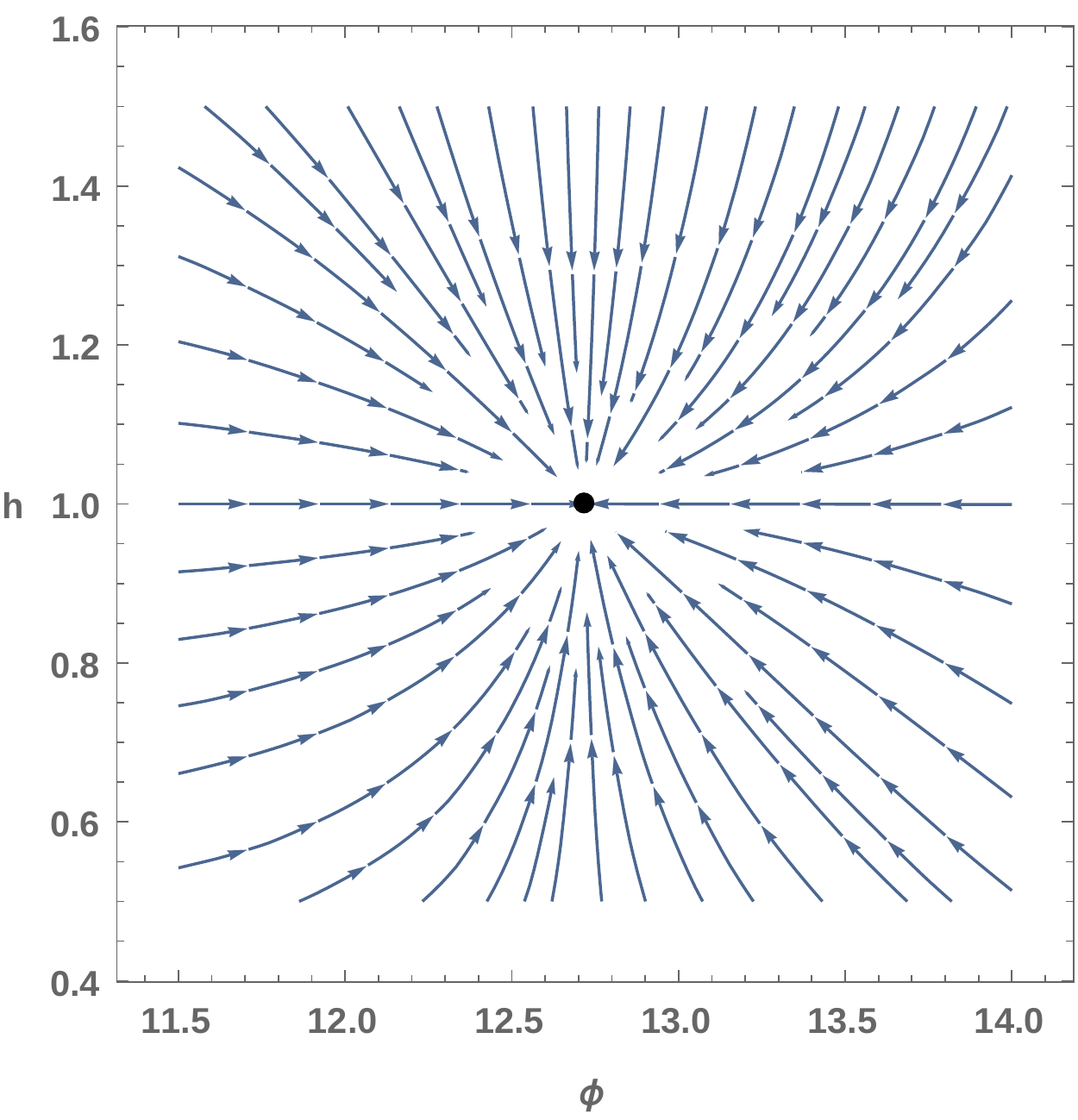}
\caption{Section of the phase space diagram for a model with $\widebar V_0=-10^{-6}\phi$ and $\widebar V_2=10^{-6}e^\phi$. The diagram depicts a section at the critical value of $\Omega_{\rm m,c}=0$.}
\label{Fig1}
\end{figure}

\begin{figure}[h]
\centering
\includegraphics[width=0.5\textwidth]{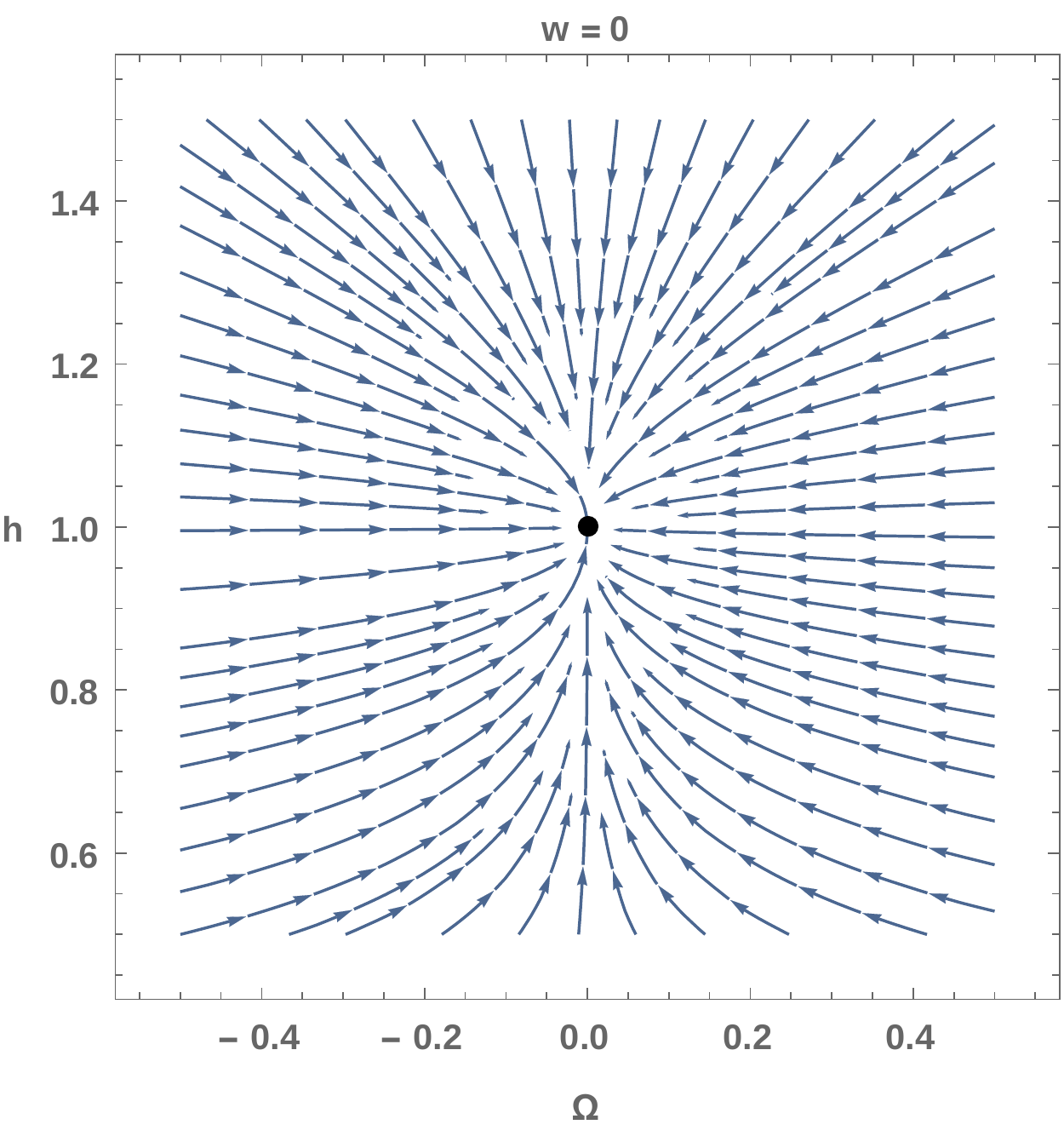}\includegraphics[width=0.5\textwidth]{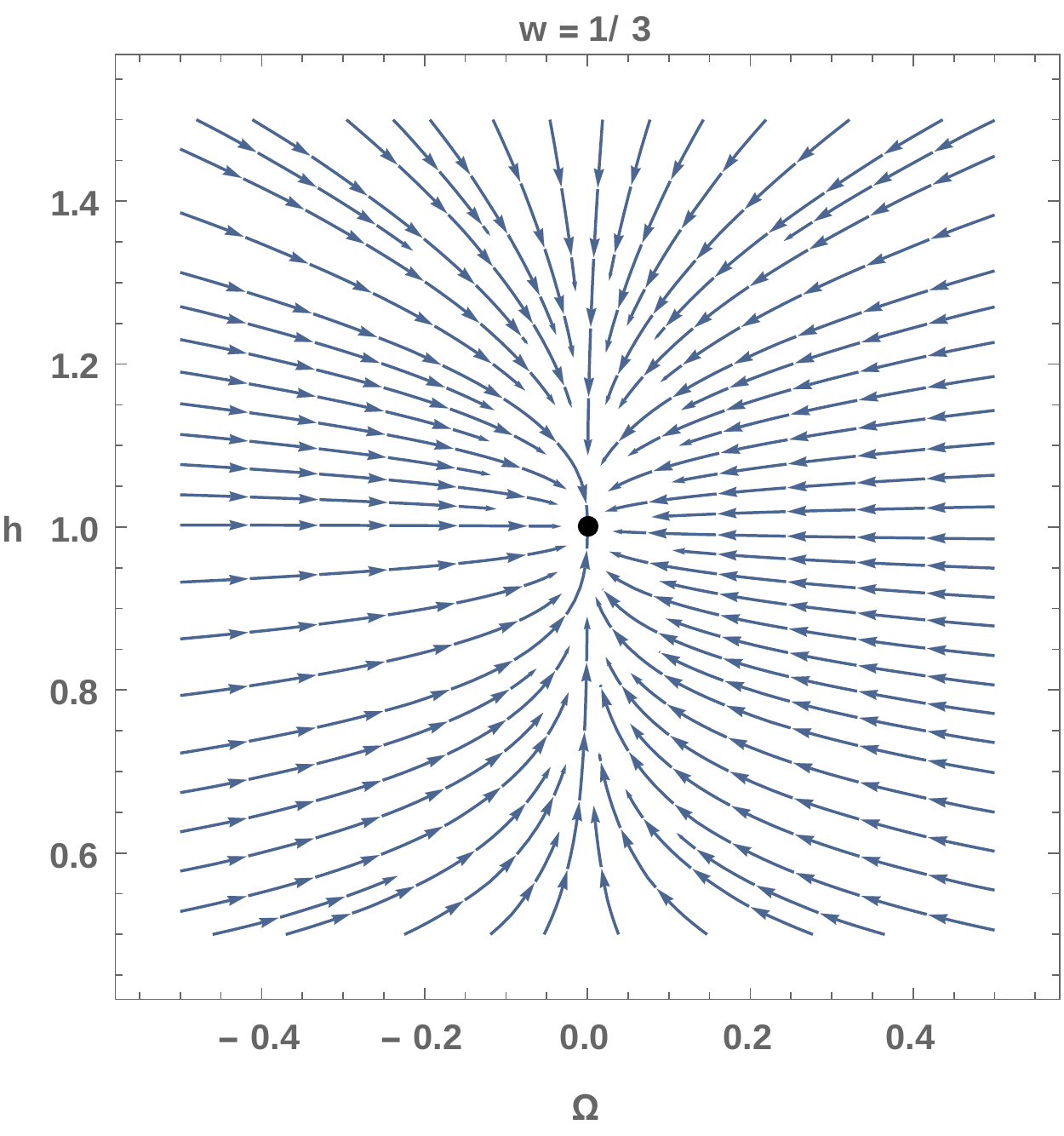}
\caption{Sections of the phase space diagram for a model with $\widebar V_0=-10^{-6}\phi$ and $\widebar V_2=10^{-6}e^\phi$.
The diagram on the left and on the right show sections at $\phi_c$ of the phase space diagram for a model only with dust matter and only with radiation, respectively.}
\label{Fig2}
\end{figure}

As we have just discussed, the dominance of the potential $V_2$ in Eq. (\ref{hs'}) in regimes with a Hubble parameter much larger than the current value, that is $h\gg1$ 
(which is equivalent to $H\gg H_0\sim\sqrt{\Lambda}$), could lead to a cosmological phase of effective matter domination, that is $ 1+w_{\rm eff}\simeq1$.
Thus, in this subsection we consider a cosmological model filled only with dust with the aim of describing the current and future evolution of our Universe.
As we are assuming dust we have
\begin{equation}
 \Omega_{\rm m}=\Omega_{{\rm m}*}\frac{h_*^2}{h^2}e^{-3\left(N-N_*\right)},
\end{equation}
where the star denotes an initial condition. 
Let us consider a model with
\begin{equation}
 \widebar V_0=-10^{-6}\phi\qquad{\rm and}\qquad \widebar V_2=10^{-6}e^\phi,
\end{equation}
and with the other potentials vanishing. Then Eqs.~(\ref{hs'}) and (\ref{phis'}) can be written as
\begin{eqnarray}
 h'&=&-\frac{3}{2}(h-1)\left(1+ \frac{e^{-\phi}}{h^2}\right), \\
 \phi'&=&\frac{(1-\Omega_{\rm m})h^2+3\times 10^{-6}\left(\phi-h^2\,e^\phi\right)}{2\times10^{-6}\,h^3\,e^\phi}.
\end{eqnarray}
Considering these expressions in Eq.~(\ref{phics}), one obtains that the model is well-defined close to the critical point, as we have a real value of the field at 
the critical point, that is, $\phi_{\rm c}=12.72$. One can explicitly check that this critical point is indeed an attractor by depicting different sections of the phase space diagram at the critical values.
In Figs.~\ref{Fig1} and \ref{Fig2}, we show that this is indeed the case. Moreover, if we consider that the universe is filled with a different fluid also with $1+w>0$, as $w=1/3$, the critical point is also an attractor
as suggested at the end of Section \ref{dynamical}.

\begin{figure}[!h]
\centering
\includegraphics[width=0.65\textwidth]{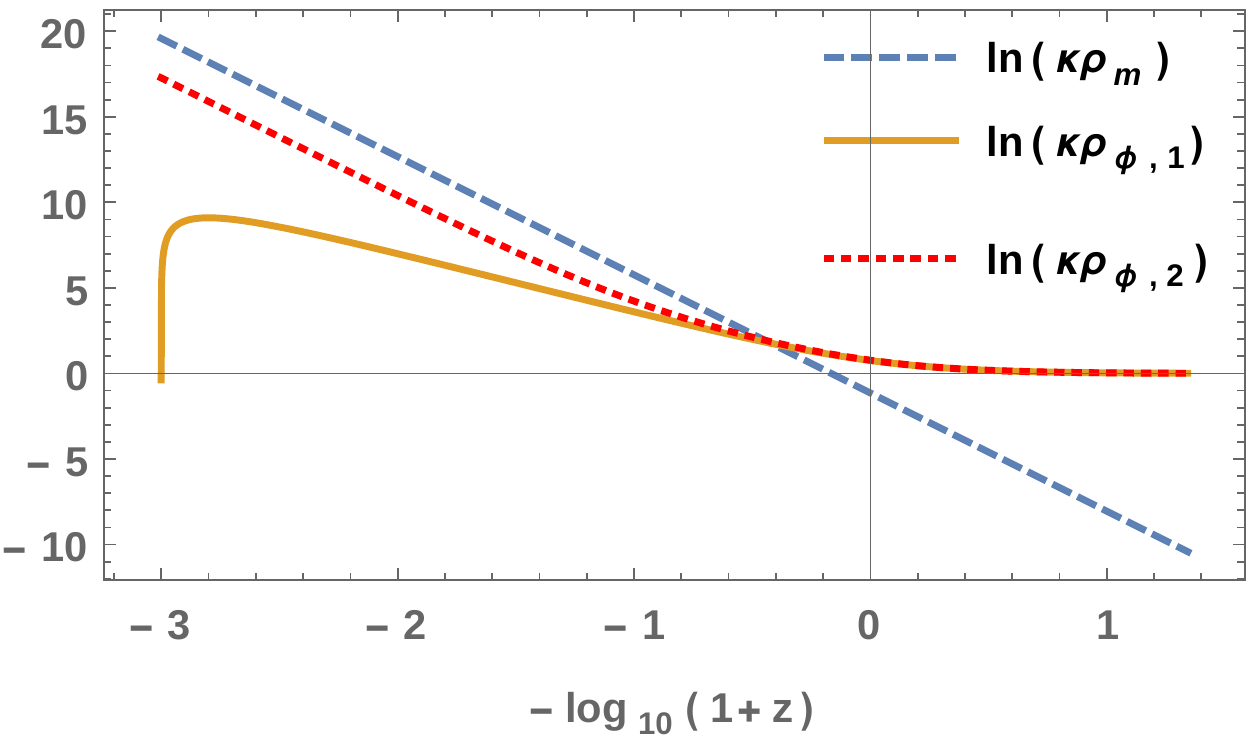}
\vspace{0.1cm}
\includegraphics[width=0.65\textwidth]{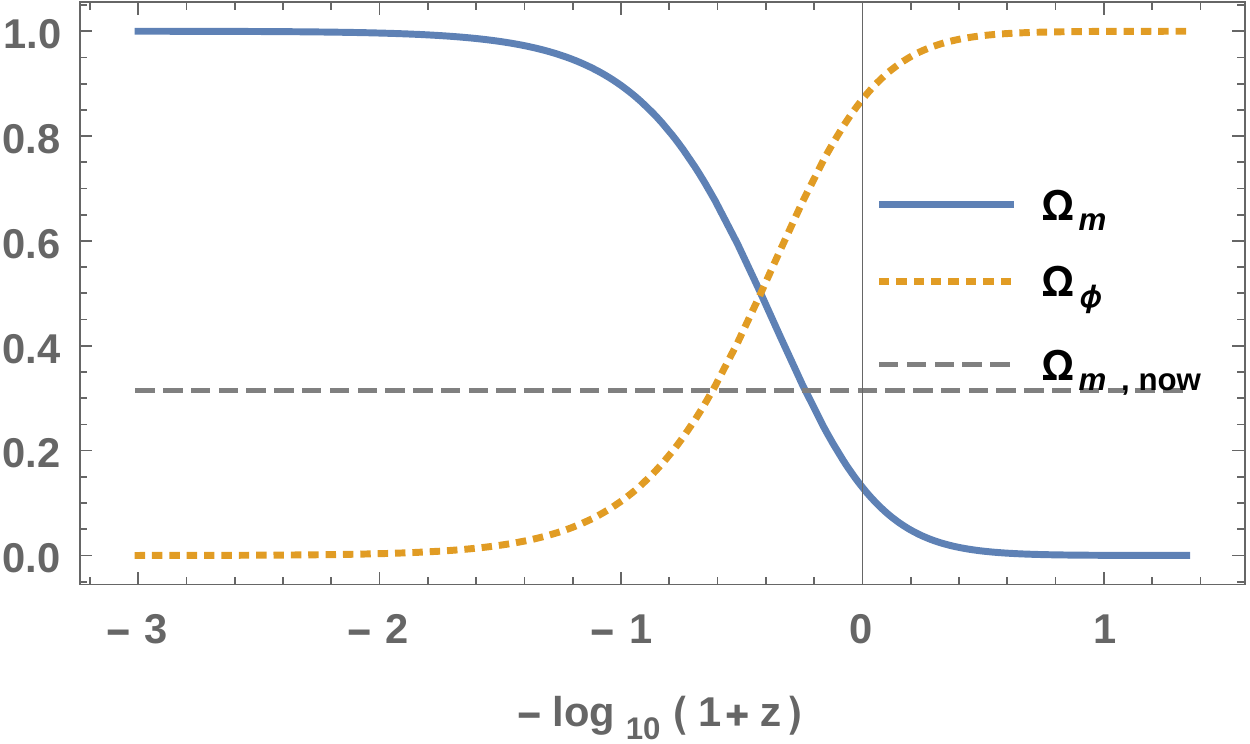}
\vspace{0.1cm}
\includegraphics[width=0.65\textwidth]{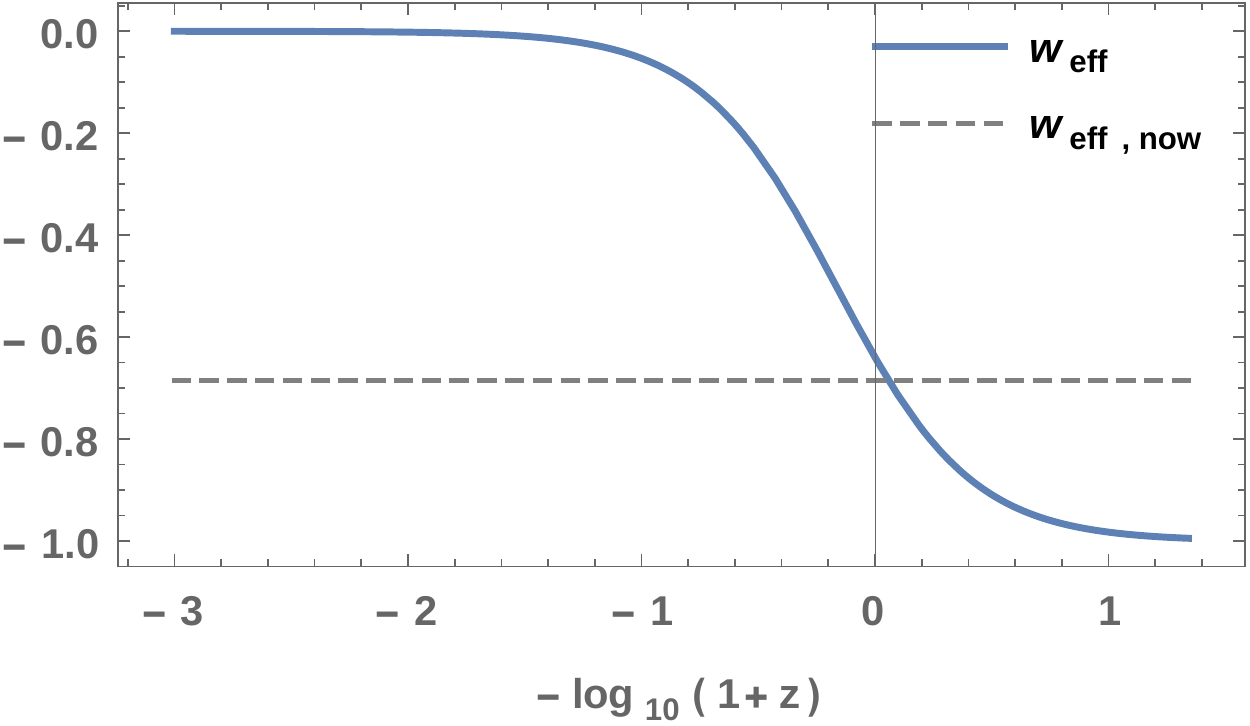}
\caption{The top panel shown the evolution of $\ln(\kappa\rho_{\rm m})$ and $\ln(\kappa\rho_\phi)$ for
a model with $\widebar V_0=-10^{-6}\phi$ and $\widebar V_2=10^{-6}e^\phi$ filled only with dust. The initial conditions used are $z_*=10^3$, $h_*=17.84\times10^3$,  
$\phi_*=1$ (a similar figure can be obtained with $\phi_* =10^3$), $\Omega_{m*}=1$ (solid line) and  $\Omega_{m*} = 0.99$ (dotted line).
The middle panel show the evolution of $\Omega_{\rm m}$ and $\Omega_{\phi}$ and the bottom panel $w_{\rm eff}$ for $\Omega_{m*}=1$.
The present values $\Omega_{{\rm m,\,now}}=0.315$ and $w_{{\rm eff,\,now}}=-0.685$ are depicted in the middle and bottom panels with dashed lines.}
\label{Fig3}
\end{figure}

One can numerically integrate the equations of the dynamics, given by Eqs.~(\ref{hs'}) and (\ref{phis'}), to represent the dynamical quantities which caracterize the model. 
In Fig.~\ref{Fig3}, we show the evolution of the matter parameter $\Omega_{\rm m}$ with respect to
\begin{equation}
\log_{10}(1+z)=-\frac{1}{\ln 10}\left[(N-N_*)-\ln\left(1+z_*\right)\right],
\end{equation}
where $z$ is the redshift, $z_*=z(N_*)$. We also compare the energy densities in logarithmic scales, 
$\ln (\kappa\rho_{\rm m})$ and $\ln (\kappa\rho_\phi)$ with $\kappa=8\pi\,G/(3 \Lambda)$.
As can be seen in Fig.~\ref{Fig3} the solution describes an early matter dominated universe that then evolves to an accelerated expansion driven by
the scalar field. Taking into account the evolution of the effective equation of state parameter given by Eq.~(\ref{weff}), we 
note that even though its value at present is within current bounds, the value of $\Omega_{\rm m}$ is smaller than the desired in order to satisfy observational constraints.   
Thus, even if the model has interesting dynamical properties, it does not entirely reproduce the cosmological evolution of our Universe. 

One could try to fine tune the parameters of the model or even change the functional form of the potentials in order to obtain a cosmological evolution closer to what observations suggest. However, this is not a trivial task because the considered model has some successful characteristics due to a long domination of $\widebar V_2$, which implies a long period of the evolution of the universe where the dynamics is simply given by Eq. (\ref{weff2}). This behaviour is also suggested by the fact that a model with 
\begin{equation}
 \widebar V_1=-10^{-6}\phi\qquad{\rm and}\qquad \widebar V_2=10^{-6}e^\phi,
\end{equation}
leads to exactly the same dynamics as the one depicted in Fig.~\ref{Fig3}. We are not suggesting that the cosmological behaviour of the models with $V_2$ and an extra $V_i$ to be trivial. For example, a model described by
\begin{equation}
 \widebar V_2=10^{-6}e^\phi\qquad{\rm and}\qquad \widebar V_3=-10^{-6}\phi,
\end{equation}
with the same initial conditions, the scalar field quickly dominates the evolution of the Universe and does not give way to a matter dominated era. This model is, 
therefore, inviable. However, a change of the initial condition for the field, for example $\phi_* = 10^3$,  affects the dynamics in such a way that one recovers the same behaviour as in Fig.~\ref{Fig3}.

\section{Tripod models}\label{tripod}

Although the models studied in the previous section can reproduce a matter dominated phase before the accelerating expansion, it is not trivial to reproduce a long enough radiation dominated epoch with those potentials.
In this section, we investigate the simplest models which can, in principle, deliver a viable cosmological history of our Universe.
To this aim we proceed as in the previous section, studying which kind of terms must dominate in order to describe this evolution. 
As these models are based on three nonvanishing potentials we call them the tripod models.

\subsection{General considerations}

Let us write schematically Eq.~(\ref{H'}) when only one $U_i$ or $W_i$ term dominates in the numerator and denominator of each of the terms. Then we have,
\begin{equation}
 H'\simeq 3\sqrt{\Lambda}\left[\frac{U_{j,\phi}H^j}{i W_{i}H^i}-\frac{ W_{i}H^{i+1}}{i W_{i}H^i}\right]\sim 3\frac{U_{j,\phi}}{i W_{i}}H^{j-i}-\frac{ 3}{i}H.
\end{equation}
Domination of the second term leads to similar conclusions as in Sec.~\ref{couple2}, that is, we can only obtain a dynamical evolution compatible with stiff matter ($i=1$), matter domination ($i=2$),
and curvature domination ($i=3$). Thus, we consider that one of the terms of our model has to be $W_2$.
As we want to describe a radiation dominated phase similar to that produced in general relativity, we  also need a regime during which the first term also dominates. Considering the first term on the r.h.s, we require $j=i+1$ such that it is proportional to $H$ (as in the dynamic equation for general relativity). In addition, the models must contain at least three potentials $U$ and $W$ 
for matter to affect the dynamics (see subsection \ref{couple}). Essentially, if there are only two potentials, we can reduce the system to a one potential case by constraint (\ref{condition}) and consequently the field dependence cancels out in the ratio of the potentials in Eq.~(\ref{H'}).
 Therefore, we consider that the simplest model must contain the potentials $W_2\neq0$, $U_{3,\phi}\neq0$ and one additional potential.
As the model with only $W_2$ and $U_2$ leads to interesting results for matter domination, we focus our attention on the model with $W_2\neq0$, $U_{2,\phi}\neq0$ and $U_{3,\phi}\neq0$.
It must be noted that this argument has helped us to find a particular working model, which does not mean that there are no other tripod models that may also be able to describe our Universe by choosing a different third potential.

Taking into account $W_2\neq0$, $U_{2,\phi}\neq0$ and $U_{3,\phi}\neq0$ in Eqs.~(\ref{H'}) and (\ref{phi'}) we obtain
\begin{eqnarray}
 h'&=&-\frac{3}{2}(h-1)\frac{\widebar U_{2,\phi}}{\widebar W_2}, \\
 \phi'&=&\frac{1-\Omega+3(2h-1)\widebar U_2-6\,h\int \widebar W_ 2{\rm d}\phi-6hc}{2\,h\,\widebar W_2},
\end{eqnarray}
where $\widebar U_2$ and $\widebar W_2$, are defined as before, and
we have considered condition (\ref{condition}) to substitute 
\begin{equation}\label{U3}
 \widebar U_{3}=\Lambda^{-1/2}\left[\int \widebar W_ 2{\rm d}\phi-\widebar U_2+c\right].
\end{equation}
The constant, $c$, specifies the particular potential $\widebar U_3$. It must be kept in mind that for the model to be well defined close to the critical point 
(when $\phi' =0$, $\Omega=\Omega_{\rm c} = 0$ and $h=h_{\rm c}=1$), the following relation
\begin{equation}
 1+3\widebar U_2-6\int \widebar W_ 2{\rm d}\phi-6c=0,
\end{equation}
must have at least one real solution $\phi_c$, as already emphasized in the general case through Eq.~(\ref{phic}).

If the de Sitter critical point is an attractor of the model and the Hubble parameter decreases towards the critical point, early enough ($h\gg 1$) one has
\begin{equation}
 h'\sim-\frac{3}{2}h\frac{\widebar U_{2,\phi}}{\widebar W_2}\qquad\Rightarrow\qquad 1+w_{\rm eff}\sim \frac{\widebar U_{2,\phi}}{\widebar W_2}.
\end{equation}
Therefore, to describe a cosmology compatible with our Universe we need the presence of potential terms such that the quotient of these two potentials
evolves from $4/3$ at early $N$ to $1$ at larger $N$. As the quotient is a function of $\phi$ we need to consider the particular material content to get the evolution in terms of $N$ as required.

\subsection{Matter and radiation with a de Sitter attractor}

Let us consider a model with
\begin{equation}\label{trippot}
 \widebar U_2=e^{\lambda\phi}+\frac{4}{3}e^{\beta\phi}\qquad{\rm and}\qquad \widebar W_2=\lambda e^{\lambda\phi}+\beta e^{\beta\phi}.
\end{equation}
$\widebar U_3$ is given by Eq.~(\ref{U3}), and the other potentials vanish. Equations (\ref{H'}) and (\ref{phi'}) can then be written as
\begin{eqnarray}
\label{triph}
 h'&=&-\frac{3}{2}(h-1)\frac{\lambda e^{\lambda\phi}+\frac{4}{3}\beta e^{\beta\phi}}{\lambda e^{\lambda\phi}+\beta e^{\beta\phi}}, \\
\label{tripphi}
 \phi'&=&\frac{1-\Omega+3\,(2h-1)\left(e^{\lambda\phi}+\frac{4}{3}e^{\beta\phi}\right)-6h\left(e^{\lambda\phi}+e^{\beta\phi}\right)-6hc}{2h\left(\lambda e^{\lambda\phi}+\beta e^{\beta\phi}\right)},
\end{eqnarray}
respectively, where $\lambda>0$ and $\beta>0$. The potentials (\ref{trippot}) have been chosen to reproduce an epoch of matter domination and another of radiation domination. This can be noted by considering the regime $h\gg1$ in Eq.~(\ref{triph}),
which implies
\begin{equation}
 1+w_{{\rm eff}}\simeq\frac{\lambda e^{\lambda\phi}+\frac{4}{3}\beta e^{\beta\phi}}{\lambda e^{\lambda\phi}+\beta e^{\beta\phi}}.
\end{equation}
Thus, one has $w_{\rm eff}\simeq0$ when the term $e^{\lambda\phi}$ dominates and $w_{\rm eff}\simeq1/3$ when $e^{\beta\phi}$ is dominant.
The actual occurrence and duration of these phases depend, however, on the behaviour of the function $\phi(N)$. If this function is monotonic, then both regimes can be attained during the evolution of the
universe. Therefore, we will choose the value of the parameters $\lambda$, $\beta$ and $c$ to ensure a positive or negative sign of $\phi'$, at least while $h\gg1$.
If we assume that the dark energy contribution is negligible when the $\beta$-term dominates, $\Omega\simeq1$, and that this happens when $h\gg1$, from Eq. (\ref{tripphi}) we have
\begin{equation}
 \phi'\simeq\frac{1-3\,c\,e^{-\beta\phi}}{\beta} \hspace{1cm}  \Rightarrow \hspace{1cm} \phi\simeq\frac{1}{\beta}{\rm ln}\left(3\,c+A\, e^N\right),
\end{equation}
In the top panel of Fig.~\ref{Fig6} we show the actual evolution of $\phi$ for a particular model.
where $A$ is an integration constant related with the initial condition.
As $\beta>0$, one would have $\phi'>0$ if $e^{\beta\phi}>3c$, and $\phi'<0$ for $e^{\beta\phi}<3c$. The first case implies that either there is a minimum value for the field, $\phi_{\rm min}$, such that $\phi > \phi_{\rm min}$, or that $c<0$, 
whereas the second case requires both $c>0$ and $\phi<\phi_{\rm max}$. Considering now the domination of the $\lambda-$term, $\Omega\simeq1$
and $h\gg1$ in Eq.~(\ref{tripphi}), one obtains
\begin{equation}
 \phi'\simeq-\frac{3\,c}{\lambda}e^{-\lambda\phi}\hspace{1cm}  \Rightarrow  \hspace{1cm} \phi\simeq\frac{1}{\lambda}{\rm ln}\left(D-3\,c\,N\right),
\end{equation}
with $D$ an integration constant. Thus, the sign of $\phi'$ is opposite to the sign of $c$. This is compatible with the considerations in the region of $\beta$-domination and does not imply 
any restriction on the value of $\phi$ if $\phi'>0$.
Additionally, the model is well behaved close to the critical point if a real $\phi_{\rm c}$ exists, that is a solution of
\begin{equation}\label{tripcrit}
 1-6c=3e^{\lambda\phi_{\rm c}}+2e^{\beta\phi_{\rm c}},
\end{equation}
by demanding $\phi'=\Omega=0$ and $h=1$ at the critical point. 
As the r.h.s. of this equation is always positive for real values of $\phi_{\rm c}$, we need to consider models with a potential $\widebar U_3$ given by Eq.~(\ref{U3}) with $c<1/6$.
Moreover, we can consider negative values of $c$ and restrict our attention to models with $\phi'(N)>0$ for $h\gg1$. In this case, one needs $\lambda>\beta$ in order to have cosmic expansion equivalent to 
radiation domination before matter domination. 
As the de Sitter critical point must be reached after the matter domination era, i.e., when the $\lambda$ term dominates over the $\beta$ term, with $\phi'=\Omega=0$ and $h=1$ in Eq.~(\ref{tripphi}), we have approximately that 
\begin{equation}
\label{phict}
\phi_{\rm c} \approx \frac{1}{\lambda} \ln\left( \frac{1-6c}{3} \right).
\end{equation}

\begin{figure}[h]
\centering
\includegraphics[width=0.5\textwidth]{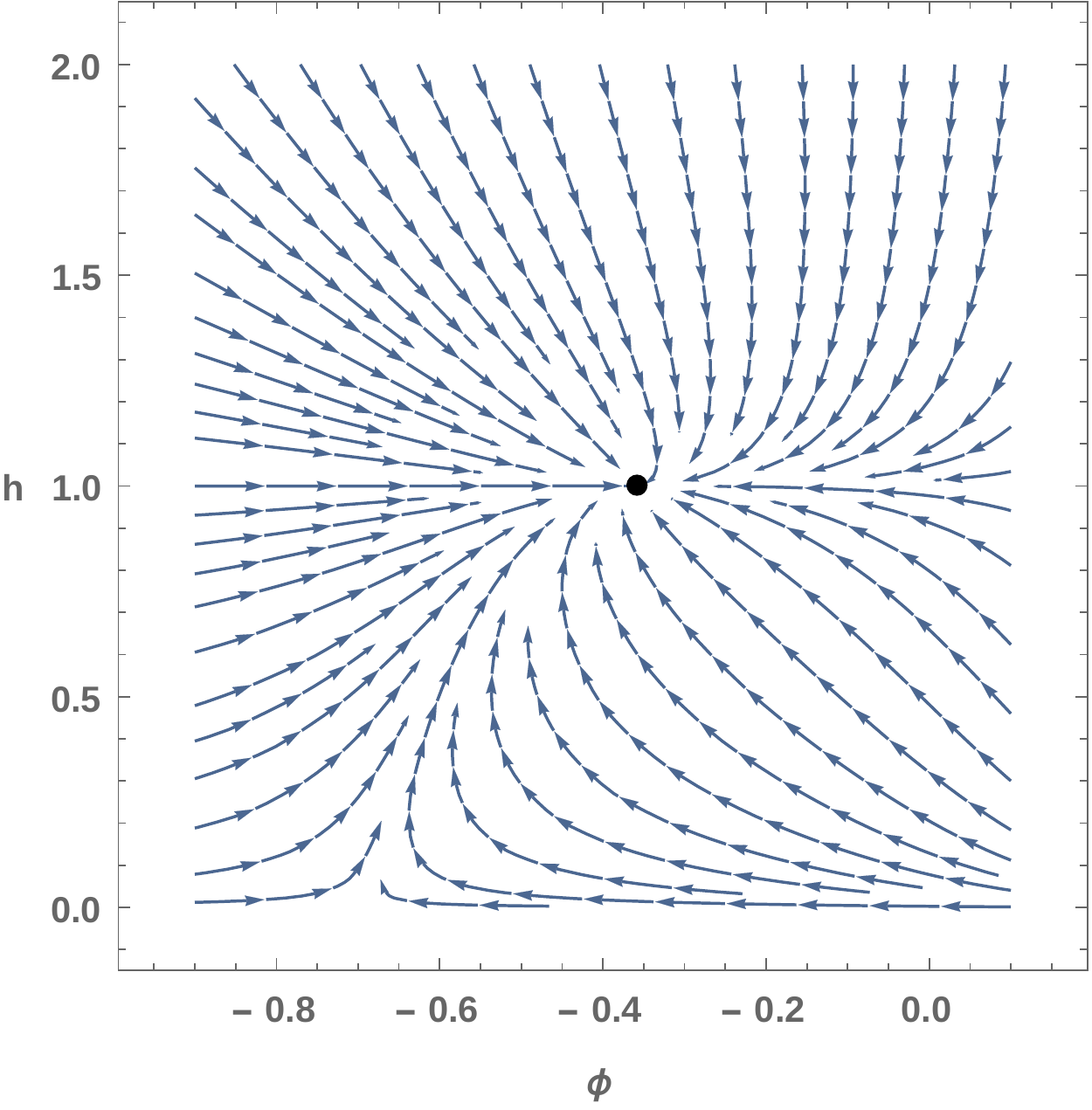}
\caption{Section of the phase space diagram for a model with $\lambda=10$, $\beta=2$ and $c=-0.01$. The diagram shows a section at the critical values $\Omega_{\rm m,c}=0$ and $\Omega_{\rm r,c}=0$.}
\label{Fig4}
\end{figure}

\begin{figure}[h]
\centering
\includegraphics[width=0.5\textwidth]{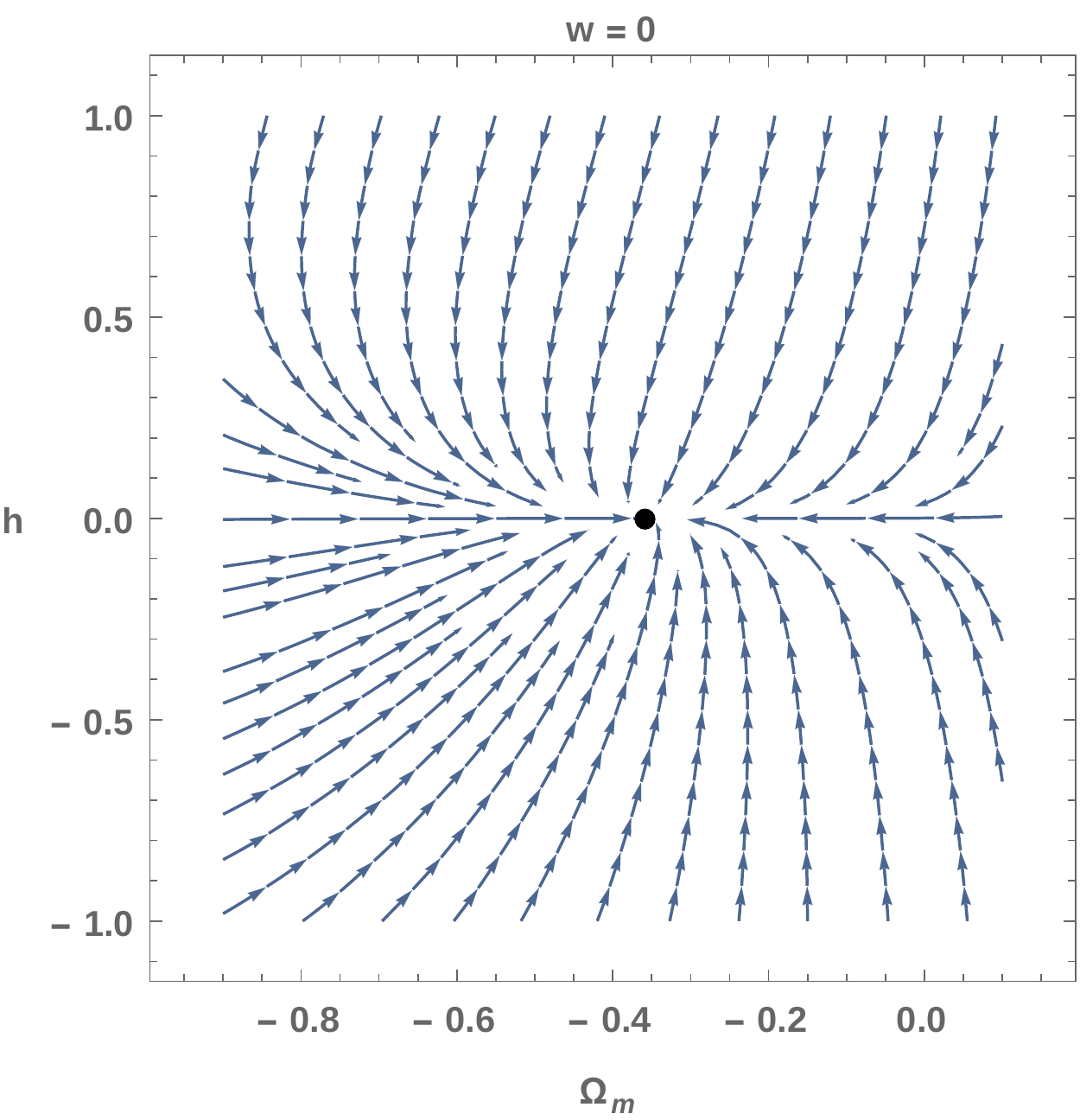}\includegraphics[width=0.5\textwidth]{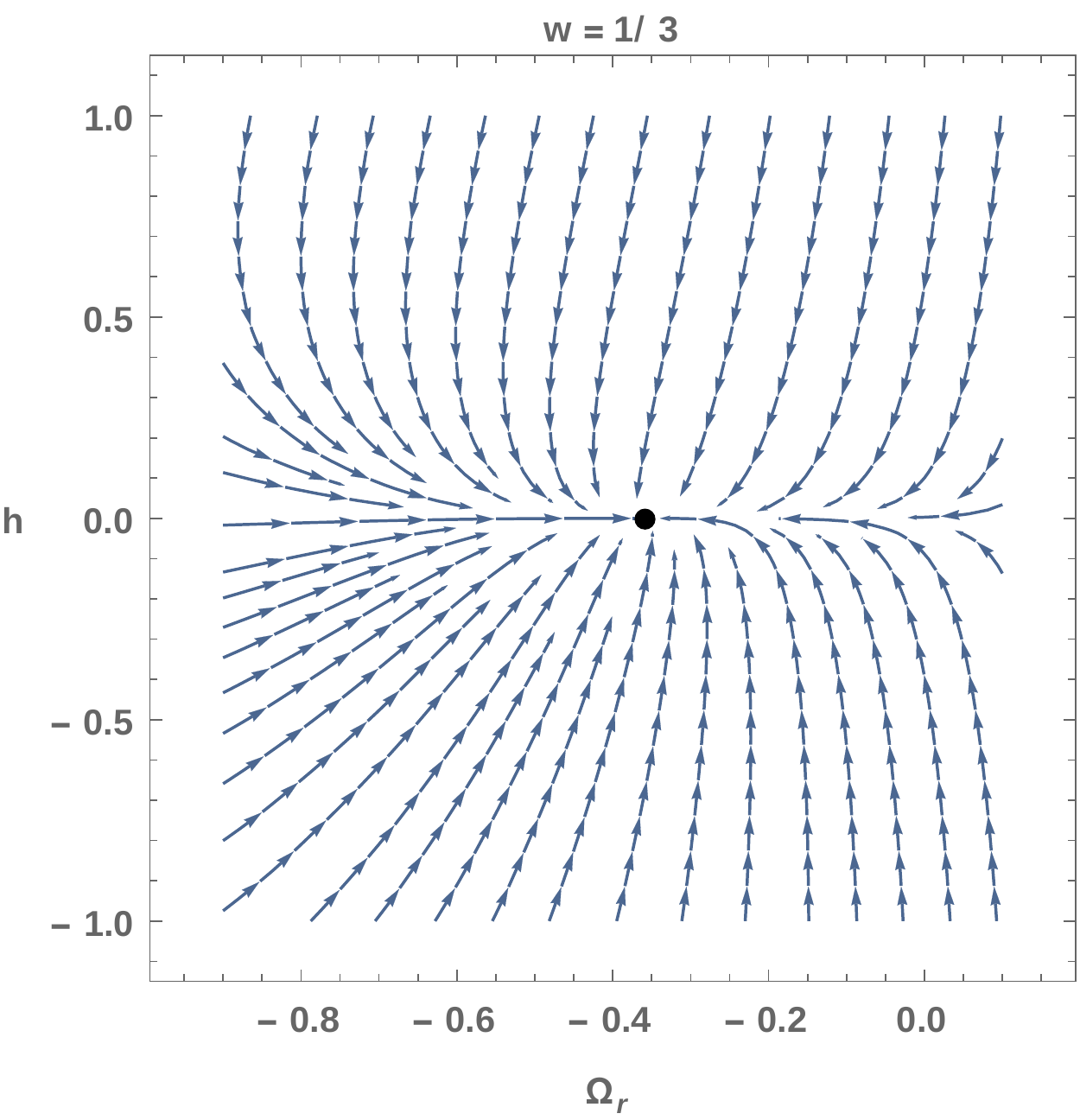}
\caption{Sections of the phase space diagram for a model with $\lambda=10$, $\beta=2$ and $c=-0.01$.
The diagram on the left shows a section at $\phi_c=-0.3584$ and  $\Omega_{\rm r,c}=0$ and that on the right is depicted using $\phi_c=-0.3584$ and  $\Omega_{\rm m,c}=0$.}
\label{Fig5}
\end{figure}

\begin{figure}[h]
\centering
\includegraphics[width=0.65\textwidth]{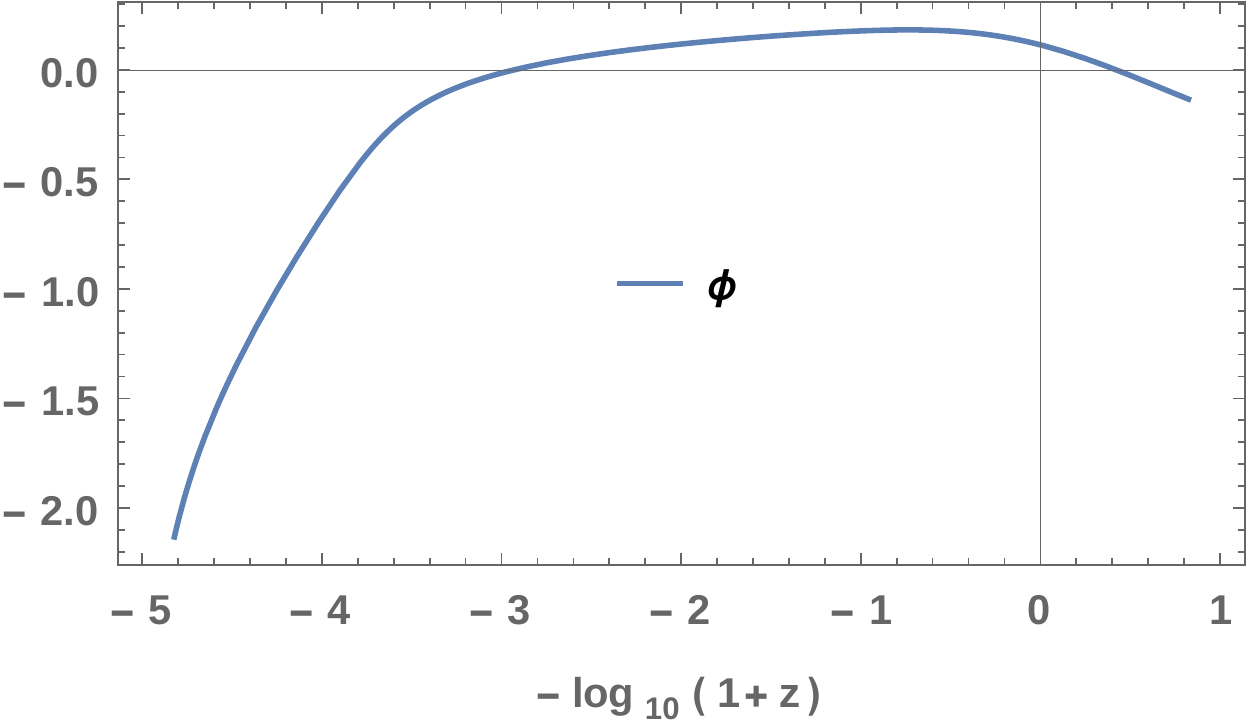}
\vspace{0.25cm}
\includegraphics[width=0.65\textwidth]{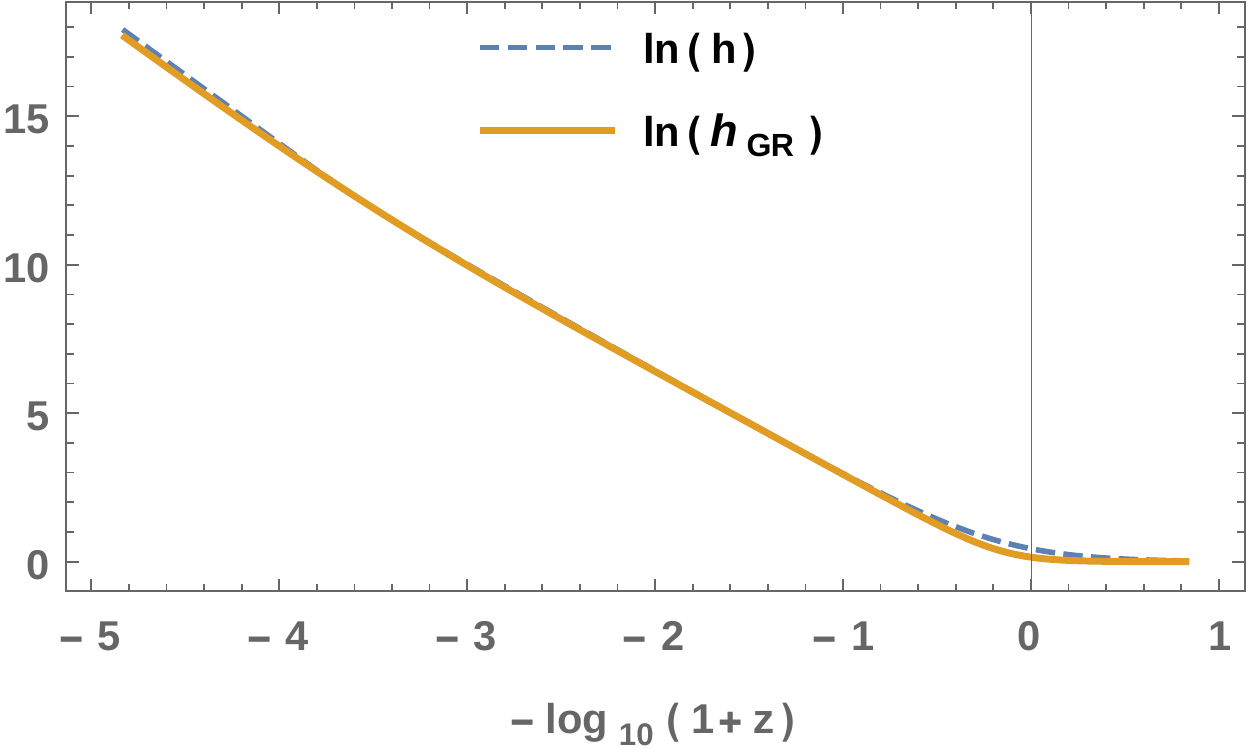}
\vspace{0.25cm}
\includegraphics[width=0.65\textwidth]{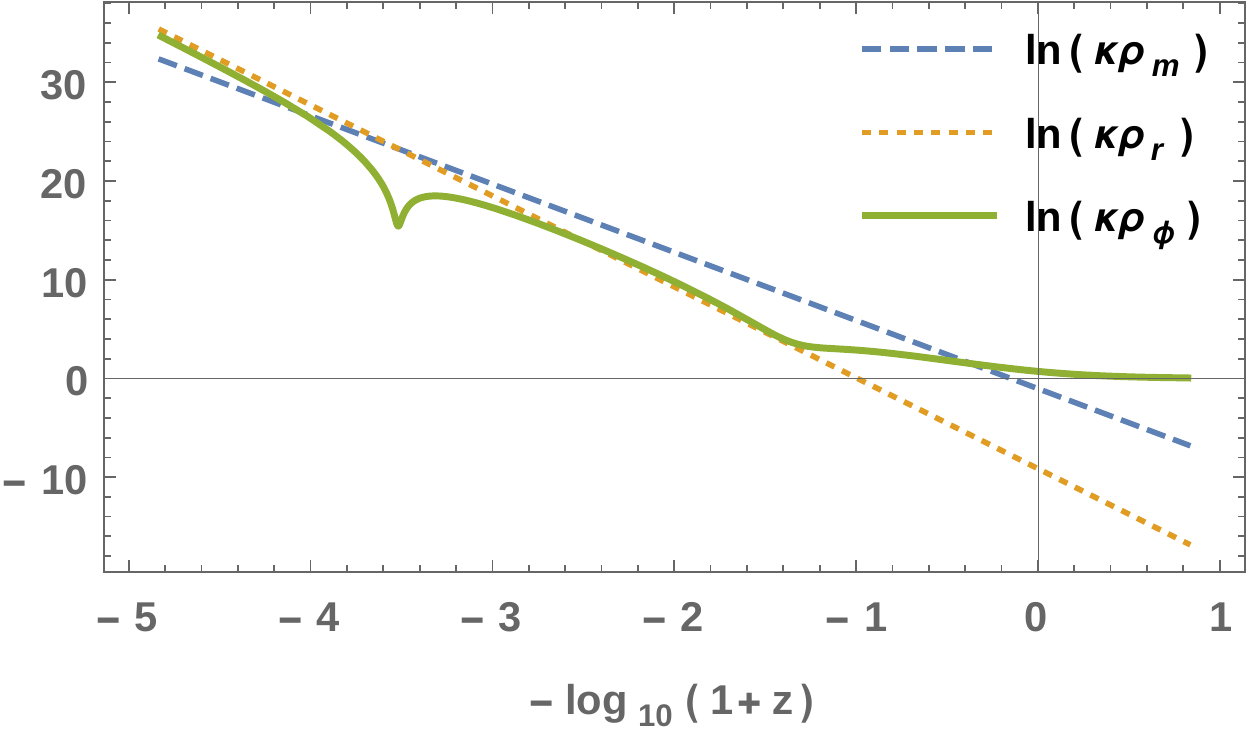}
\caption{Model with $\lambda=10$, $\beta=2$ and $c=-0.01$, and conditions imposed at matter-radiation equivalence, 
$z_{{\rm eq}}=3.3\times10^3$, $\Omega_{\rm m,eq}=\Omega_{\rm r,eq}=0.4999$, $h_{\rm eq}=1.6\times10^5$,
and $\phi_{\rm eq}=-0.20118$.
On the top we show the evolution of the scalar field.
In the centre we compare the normalized Hubble parameter of our model with that of a $\Lambda$CDMR model with the same conditions at $N_{\rm eq}$ using logarithmic scales. 
At the bottom we show the evolution of the energy density of matter, radiation and the dark energy component.}
\label{Fig6}
\end{figure}

In Figs.~\ref{Fig4} and \ref{Fig5}, we show the phase space diagram for a particular choice of the parameters of the model. We are considering both matter and radiation, therefore, the phase space is 4-dimensional 
and the critical point is characterized by
\begin{equation}
 \{h_{\rm c}=1, \, \Omega_{\rm m,c}=0, \, \Omega_{\rm r,c}=0, \, \phi_{\rm c}\},
\end{equation}
with $\phi_{\rm c}$ given by Eq.~(\ref{tripcrit}).
Thus, in order to understand the attractor nature of the critical point, we have depicted different sections of the phase space diagram.

\begin{figure}[h]
\centering
\includegraphics[width=0.7\textwidth]{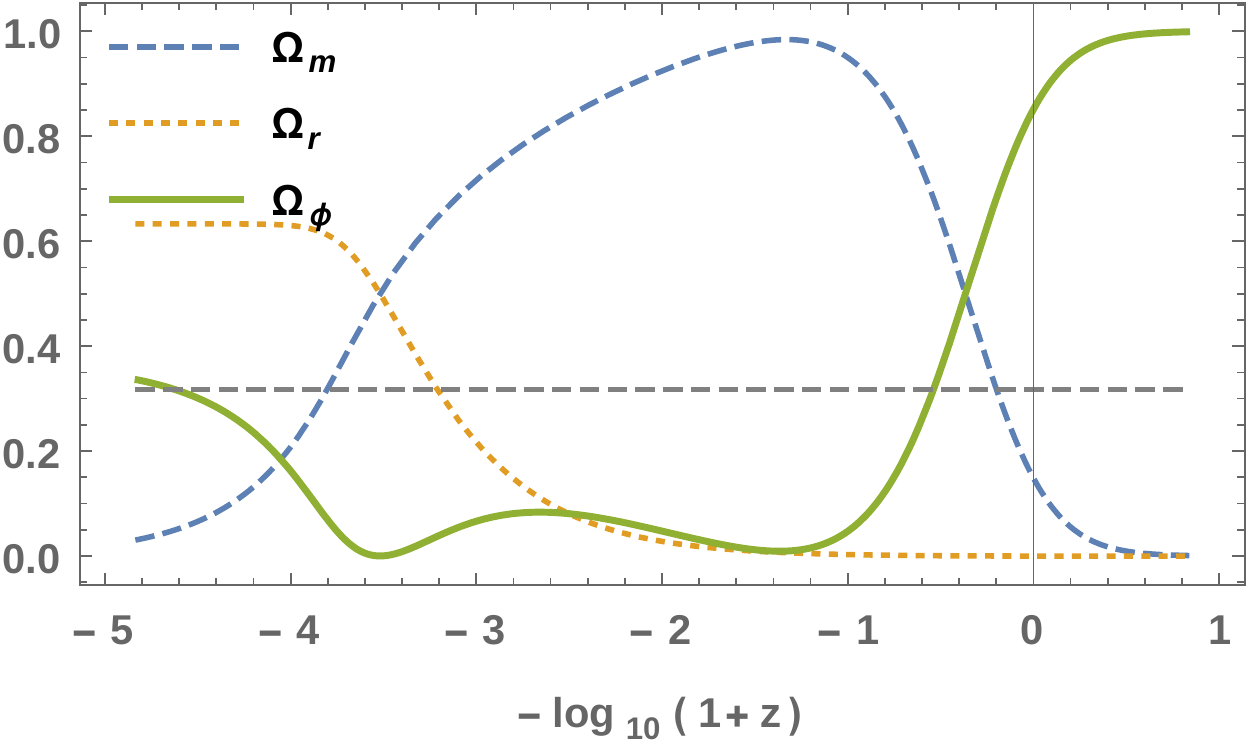}
\vspace{0.25cm}
\includegraphics[width=0.7\textwidth]{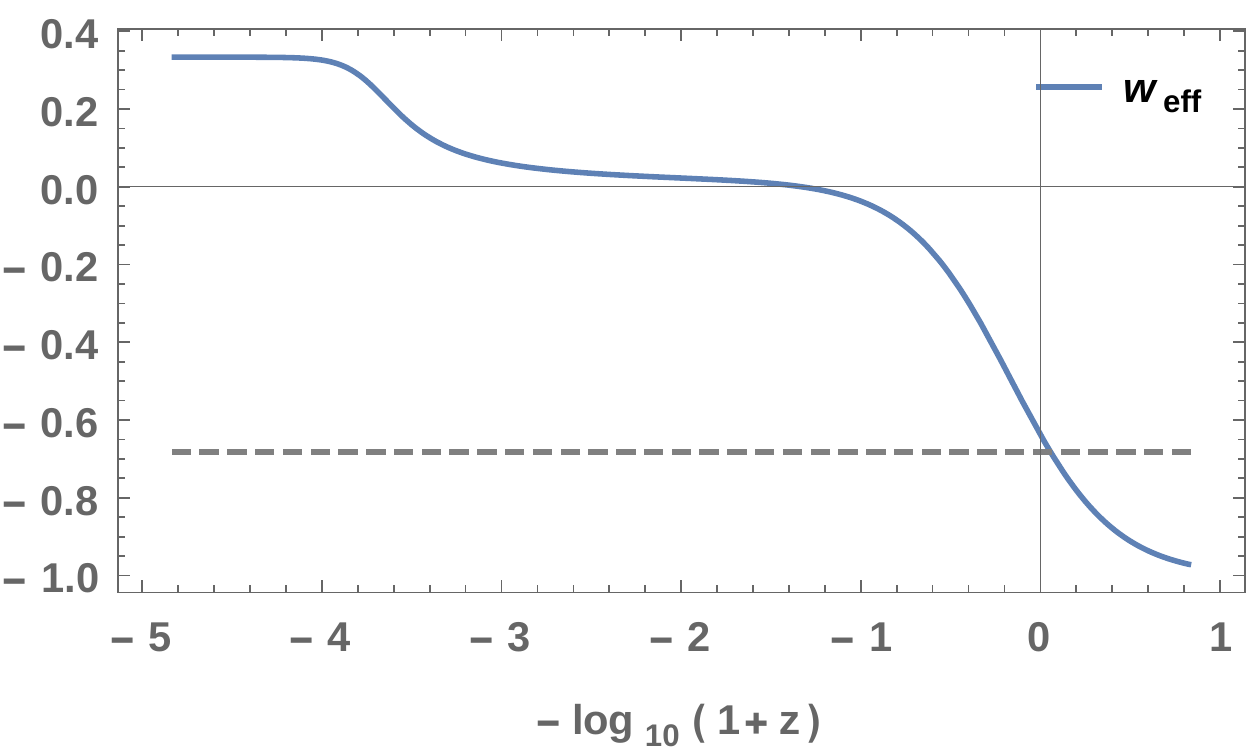}
\caption{Model with $\lambda=10$, $\beta=2$ and $c=-0.01$, and conditions imposed at matter-radiation equivalence, $\Omega_{\rm m,eq}=\Omega_{\rm r,eq}=0.4999$, $h_{\rm eq}=1.6\times10^5$,
and $\phi_{\rm eq}=-0.20118$. 
We show, on the top panel, the evolution of the $\Omega$ parameters of the model and, on the bottom panel, the effective equation of state parameter. In dashed lines we include the observationally measured values of $\Omega_{\rm m,now}=0.3175$ and
$w_{\rm eff,now}=-0.6825$.}
\label{Fig7}
\end{figure}

One can numerically integrate the equations of the dynamical system (\ref{triph}) and (\ref{tripphi}) considering a universe filled with non-relativistic matter and radiation. Assuming that this happens at large enough values of $h$
and imposing the general relativistic value $w_{\rm eff,eq}=1/6$, one can obtain the condition for the field at the matter-radiation equivalence, which is given by
\begin{equation}
\label{phieq}
 \phi_{\rm eq}\simeq\frac{1}{\lambda-\beta}{\rm ln}\frac{\beta}{\lambda}.
\end{equation}
In Figs.~\ref{Fig6} and \ref{Fig7} we show the dynamical quantities obtained via this integration in terms of $-{\rm log}_{10}(1+z)=\left[(N-N_{\rm eq})-{\rm ln}\left(1+z_{\rm eq}\right)\right]/{\rm ln}10$.
The central panel of Fig.~\ref{Fig6} depicts the comparison of the evolution of the Hubble parameter of our model with that of a $\Lambda$CDMR model (cosmological constant, cold dark matter and radiation) in logarithmic scales. Note that our model can reproduce the Hubble parameter of the mentioned general relativistic model with great accuracy. As the luminosity distance is calculated integrating $H(z)$, this result suggests that our model could equally reproduce the expected form of this distance. 
Moreover, the bottom panel of Fig.~\ref{Fig6} depicts the evolution of the different energy densities in logarithmic scale. The dark energy component tracks radiation during radiation and matter domination until it approaches the constant value of the critical point.

The top panel of Fig.~\ref{Fig7} depicts the evolution of the $\Omega_s$ parameters. The qualitative behaviour of these parameters is consistent with a viable cosmological evolution. 
There is indeed a radiation dominated era followed by a matter dominated epoch and only at low redshift the scalar field takes over the evolution leading the Universe to a de Sitter stage. 
Quantitatively we cannot help noticing that the contribution of the scalar field at recombination is certainly larger than allowed by current limits which imply at least $\Omega_\phi<0.02$ \cite{Ade:2015rim}.
In the bottom panel of Fig.~\ref{Fig7} we show the evolution of the effective equation of state parameter given by Eq.~(\ref{weff}). From Fig.~\ref{Fig7} we can understand that even though the effective equation of state parameter is within observational bounds, the abundance of the scalar field at present, however, is too large.
Tuning the parameters $\lambda$, $\beta$ and $c$, does not improve the dynamics in a substancial form and we can only conclude that  even though this model presents the adequate qualitative features, it does not pass the observational constraints. 

\section{Summary and further comments}\label{discussion}

In this article we have considered the cosmology of the family of models linear in the field derivatives of the most general minisuperspace Lagrangian for a scalar field leading to field equations with no higher than second order derivatives and self tuning to a spatially flat de Sitter vacuum. We have studied the dynamical system of these cosmologies concluding that the spatially flat de Sitter critical point is not hyperbolic and, therefore, one has to consider particular models to conclude whether this point is an attractor going beyond the linear analysis.
We have shown that, due to the particular form of the Lagrangian the material content affects the cosmological dynamics through its effect on the scalar field. We have studied some simple models for which, however, matter does not matter, since the differential equation for the Hubble parameter is independent of both the material content and of the field value. These toy models are of special interest for understanding the behaviour of the different potentials when they have the dominant contribution in more complex models.

We have then moved to consider particular models. In the first place, we considered what we called ``term-by-term'' (TBT) models, since they satisfy the constraint for the potentials term by term. As we have seen, 
it is not difficult to describe an early matter dominated universe that then evolves to an accelerated expansion driven by the scalar field. Nevertheless, at least for the particular choice of the potentials that 
we considered, the material content decays slightly faster than expected. It must be noted, however, that by no means the mechanism is so efficient to immediately screen any material content, as suggested in 
Ref.~\cite{Linder:2013zoa}  concerning the models presented in Ref.~\cite{Appleby:2012rx}. In the second place, we investigated the tripod models, where only three potentials are present, as these models can be 
the simplest cases capable of producing a satisfactory cosmological history. These models are able to describe a radiation dominated epoch preceding matter domination and to have a de Sitter attractor, although 
the material content also decays slightly faster than expected in the particular case that we have considered in more detail.  Therefore, even though 
we have not found an optimal example, the qualitative good features of these 
models render them promising candidates to describe our Universe's evolution.

From the purely formal point of view it is worthy to note that the models considered in this article may avoid the cosmological constant problem at the cost of requiring fine tuning of a number of parameters appearing in the specific models. The cosmological constant problem, however, is not solely based on the fine tuning of a constant of nature when it is understood as vacuum energy, but in the necessity of fine tuning it depending on the phase transitions that may lead to different vacuum energies. Thus, the problem could be thought to be alleviated since the fine tuning can now be undertaken independently of the underlying particle physics theory describing the Universe.

\begin{acknowledgments}
The authors acknowledge financial support of the Funda\c{c}\~{a}o para a Ci\^{e}ncia e Tecnologia through the grant EXPL/FIS-AST/1608/2013. 
PMM is also suppported by the grant PTDC/FIS/111032/2009, and
NJN by OE/FIS/UI2751/2014. FSNL acknowledges financial support by an Investigador FCT Research contract, with reference IF/00859/2012.
\end{acknowledgments}


\end{document}